\documentclass[twocolumn,aps,pra,superscriptaddress,floatfix]{revtex4}
\usepackage{graphicx}  % needed for figures
\usepackage{dcolumn}   % needed for some tables
\usepackage{bm}        % for math
\usepackage{amssymb,amsmath}        % for math
\usepackage[usenames]{color}
\usepackage[colorlinks,urlcolor=blue,citecolor=blue,linkcolor=blue]{hyperref}

\begin{document}

\newcommand{\bra}[1]{\mbox{\ensuremath{\langle #1 \vert}}}
\newcommand{\ket}[1]{\mbox{\ensuremath{\vert #1 \rangle}}}
\newcommand{\mb}[1]{\mathbf{#1}}
\newcommand{\phipp}{\big|\phi_{\mb{p}}^{(+)}\big>}
\newcommand{\phipav}{\big|\phi_{\mb{p}}^{\p{av}}\big>}
\newcommand{\pp}[1]{\big|\psi_{p}(#1)\big>}
\newcommand{\drdy}[1]{\sqrt{-R'(#1)}}
\newcommand{\Rb}{$^{87}$Rb\ }
\newcommand{\kf}{$^{40}$K}
\newcommand{\na}{${^{23}}$Na}
\newcommand{\muK}{\:\mu\textrm{K}}
\newcommand{\p}[1]{\textrm{#1}}
\newcommand\T{\rule{0pt}{2.6ex}}
\newcommand\B{\rule[-1.2ex]{0pt}{0pt}}
\newcommand{\reffig}[1]{\mbox{Fig.~\ref{#1}}}
\newcommand{\refeq}[1]{\mbox{Eq.~(\ref{#1})}}
\hyphenation{Fesh-bach}
\newcommand{\previous}[1]{}
\newcommand{\note}[1]{\textcolor{red}{[\textrm{#1}]}}
\newcommand{\nuke}[1]{}

\newcommand{\pfreq}{\ensuremath{\Omega}}
\newcommand{\ifreq}{\ensuremath{\Phi}}
\newcommand{\rperp}{\ensuremath{R_\perp}}

%\title{The Vortex Spirograph: Dipole Dynamics in Bose-Einstein Condensates}
\title{Guiding-center dynamics of vortex dipoles in Bose-Einstein condensates}

\author{S.\ Middelkamp}
\affiliation{ Zentrum f\"ur Optische Quantentechnologien, Universit\"at
Hamburg, Luruper Chaussee 149, 22761 Hamburg, Germany}
\author{P.\,J.\ Torres}
\affiliation{Departamento de Matem\'atica Aplicada,
Universidad de Granada, 18071 Granada, Spain}
\author{P.\,G.\ Kevrekidis}
\affiliation{Department of Mathematics and Statistics, University of Massachusetts,
Amherst MA 01003-4515, USA}
\author{D.\,J.\ Frantzeskakis}
\affiliation{Department of Physics, University of Athens, Panepistimiopolis,
Zografos, Athens 157 84, Greece}
\author{R.\ Carretero-Gonz\'{a}lez}
\affiliation{
Nonlinear Dynamical System Group,
%({\tt http://nlds.sdsu.edu}),
Computational Science Research Center, and
Department of Mathematics and Statistics,
San Diego State University, San Diego, California 92182-7720, USA}
\author{P.\ Schmelcher}
\affiliation{ Zentrum f\"ur Optische Quantentechnologien, Universit\"at
Hamburg, Luruper Chaussee 149, 22761 Hamburg, Germany}
\author{D.\,V.\ Freilich}
\affiliation{Department of Physics, Amherst College, Amherst, Massachusetts, 01002-5000 USA}
\author{D.\,S.\ Hall}
\affiliation{Department of Physics, Amherst College, Amherst, Massachusetts, 01002-5000 USA}
%}

\begin{abstract}
A quantized vortex dipole is the simplest vortex molecule, comprising two counter-circulating vortex lines
%embedded within
in a superfluid. Although vortex dipoles are endemic in two-dimensional superfluids,
%and form the elementary building block of quantum turbulence; yet
the precise details of their dynamics have remained largely unexplored. We present here several striking observations of vortex dipoles in dilute-gas Bose-Einstein condensates, and develop a
%consider a lowest-order
vortex-particle model that generates vortex line trajectories that are in good agreement with the experimental data.
%Surprisingly
Interestingly, these diverse trajectories exhibit essentially identical quasi-periodic behavior, in which the vortex
lines undergo stable epicyclic orbits.
\end{abstract}

\maketitle

{\it Introduction.---} Vortices are persistent circulating flow patterns that occur in an extraordinary variety of scientific and mathematical contexts \cite{Lugt1983,Pismen1999}. In superfluids
and superconductors, quantized vortices are topological excited collective states that play central roles in transport and dissipative properties.
%The realization of
Dilute-gas Bose-Einstein condensates~\cite{Pethick2002,Pitaevskii2003} have supplied a particularly pristine setting for the %examination
study of static and dynamic vortex configurations, ranging from solitary vortex lines to large vortex lattices~\cite{Fetter2001,Fetter2009}. Our primary focus in this Rapid Communication is on the behavior of the vortex dipole ---a self-assembled, counter-circulating pair of quantized vortices--- which lies at the heart of phenomena associated with quantum turbulence~\cite{Feynman1955} and two-dimensional (2D) degenerate Bose gases~\cite{Hadzibabic2006}. Only recently have vortex dipoles been realized experimentally in dilute-gas Bose-Einstein condensates~\cite{Dutton2001,Neely2010,Freilich2010}, and no unifying theoretical perspective on their dynamical behavior has yet emerged.

Vortex dipoles in weakly-interacting superfluid systems were first considered as one of several stationary configurations of vortex clusters occurring in harmonically confined, quasi-2D
%two-dimensional
geometries~\cite{Crasovan2002,Crasovan2003}. Subsequent theoretical studies determined their energy and angular momentum~\cite{Zhou2004},
and two independent modes were identified
under small perturbations: a zero-frequency, precessional mode~\cite{Pietila2006}, and an oscillatory mode about the stationary fixed points~\cite{Li2008}.
%We note that
Additional stationary vortex cluster configurations, such as
%have since been predicted, including
vortex tripoles~\cite{Mottonen2005} and
%a variety of
other, more exotic arrangements~\cite{Middelkamp2010} have also been predicted.

Experiments involving vortex clusters are of more recent vintage~\cite{Neely2010,Seman2010,Freilich2010}.
Quantized vortex dipoles and the trajectories of their individual vortex lines were first observed experimentally by translating a repulsive obstacle through a Bose-Einstein condensate~\cite{Neely2010}; these observations were accompanied by an %three-dimensional
integration of the three-dimensional (3D)
%mean-field model, the
Gross-Pitaevskii (GP) equation~\cite{Pethick2002}, which achieved good qualitative agreement with the experimental trajectories of the individual vortex lines. The development of a real-time vortex imaging method led to the subsequent identification of asymmetric and stationary symmetric vortex dipole configurations~\cite{Freilich2010}. A recent theoretical analysis~\cite{Kuopanportti2011} of these results found good agreement with the static properties of the symmetric dipole, but met with less success in explaining the dynamics of the asymmetric dipoles when using the 3D GP model.
%Gross-Pitaevskii model.
%a two-dimensional reduction of the Gross-Pitaevskii equation.

We present here a detailed experimental study of vortex dipole dynamics in trapped Bose condensates, and we develop a theoretical framework that discerns the fundamental patterns of vortex motion by concentrating directly on the vortices themselves.
%The theory leads to a surprising result:
We find that, to lowest order, the vortices in a dipole behave as classical particles, exhibiting quasi-periodic dynamics in which they execute stable, cyclic orbits about rigidly precessing guiding centers.
%We further establish the conserved quantities of the system and demonstrate its stability and integrability.
Ultimately, we achieve an intuitive understanding of the underlying dynamics, in which the compelling agreement between theory and experimental results reveals that \emph{all} such vortex dipoles exhibit essentially identical quasi-periodic behavior.

{\it Experiment.---} Our apparatus and experimental procedure are described elsewhere~\cite{Mertes2007,Freilich2010}. We begin with an oblate \Rb condensate of $\sim 6\times 10^5$\,atoms in the $\ket{F=1, m_F=-1}$ state, confined in a TOP magnetic trap~\cite{Petrich1995} with effective frequencies $\{\omega_r,\omega_z\} = 2\pi \cdot \{35.8(2), 101.2(5)\}$\,Hz. The Thomas-Fermi radius of the condensate is $\rperp \sim 17~\mu$m. Vortex dipoles arise spontaneously~\cite{Weiler2008} but infrequently ($\sim 3$\% of the time) during the final stage of evaporative cooling. To observe the vortex dynamics we use a real-time imaging method~\cite{Freilich2010} that repeatedly extracts, expands, and absorptively images small (1\% or 5\%) fractions of the atomic sample. A final time-of-flight image of the remaining atoms is taken at the end of the experiment.

\begin{figure}[ht]
\includegraphics[width=7.0cm]{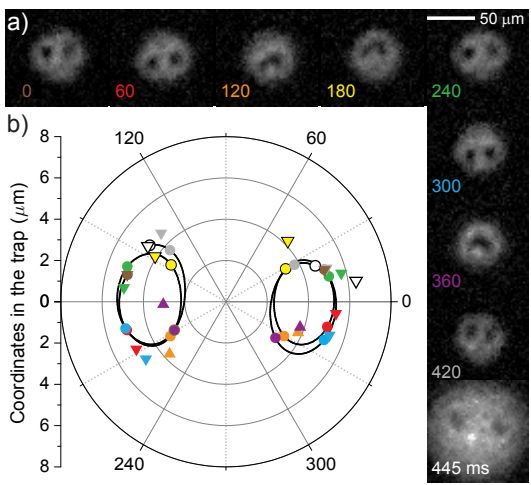}
\caption{\label{fig:epicyclic}
(color) Symmetric vortex configuration.
(a) Experimental images of the atomic density (optical depth), 1\% extraction.
(b) Polar plot of the experimental data (triangles) and the theoretical prediction from \refeq{eq:ode}
(solid line). The colored circles correspond to the calculated vortex coordinates at the same times as the experimental data. Downward (upward) triangles indicate configurations in which a surface fit to the images in (a) succeeded (failed) to distinguish the vortices. Only the inner region of the condensate is shown.
}
\end{figure}

Image sequences of the motion of the vortex lines are shown in Figs.~\ref{fig:epicyclic} and~\ref{fig:translated}. A surface fit to each image determines the coordinates of each vortex and the condensate center. Vortex coordinates  that cannot be fit are approximated by direct inspection. We destructively measure the initial number of atoms in a typical condensate at the end of the evaporation cycle, and measure the final number of atoms in subsequent condensates that bear vortex dipoles. We estimate the number of atoms at intermediate times based on the fraction extracted for each image, including an extra reduction (typically 1\%) between images to reflect atomic losses due to three-body recombinations~\cite{Burt1997}.

The sudden reduction in the number of atoms following expansion is expected to excite a small-amplitude (a few percent) breathing mode, with only slight alteration of vortex trajectories~\cite{Kuopanportti2011}. Experimentally, we discern neither the anticipated excitation nor any disturbance of the motion of a solitary vortex line~\cite{Freilich2010}, further suggesting that any effect on a dipole is small. A full exploration of the imaging process and its effect on dipole dynamics
awaits a theory of extraction and is therefore beyond the scope of the present work.

{\it Theory.---} We adopt an effective particle model in order to understand and theoretically follow the fundamental characteristics of the observed vortex dynamics and interactions. This type of modeling has been shown to be successful in describing single vortex dynamics~\cite{Svidzinsky2000,Fetter2001} and, more recently, in capturing some multi-vortex features~\cite{Schulte2002,Li2008,Middelkamp2010}. Particle modeling has also proven useful in identifying equilibrium states, oscillatory modes, and far-from-equilibrium collisional dynamics of matter-wave dark solitons
%(the one-dimensional analogs of the vortices) in Bose-Einstein condensates~
\cite{Weller2008,Theocharis2010,Coles2010}.

The model for a vortex pair involves only two ingredients, and is readily generalized to $n$ vortices~\cite{Middelkamp2010}. The first ingredient is the gyroscopic precession of a solitary vortex line induced by the inhomogeneous atomic density profile of the condensate. In the absence of dissipation, this amounts to uniform circular motion at a fixed distance $r$ from the condensate center. For a nonrotating, axisymmetric, disk-shaped trap, the angular frequency $\pfreq(r)$ of the precession is given by \cite{Fetter2001,Freilich2010}:
%(as found theoretically~\cite{Fetter2001} and supported experimentally~\cite{Freilich2010}):
%
\begin{equation}\label{eq:omegar}
\pfreq(r) = \frac{2\hbar\omega_r^2}{8\mu(1-r^2/\rperp^2)}\left(3 + \frac{\omega_r^2}{5 \omega_z^2}\right)\ln\left(\frac{2\mu}{\hbar \omega_r}\right),
\end{equation}
where $\omega_r$ ($\omega_z$) is the radial (axial) trap frequency, $\mu$ is the chemical potential, and $\rperp$ is the Thomas-Fermi radius.
%106
%
The second ingredient is the pairwise interaction between the vortex lines. For two vortex lines (labeled by subscripts $j$ and $k$) in a \emph{homogeneous} condensate, each point on one vortex line moves in the direction of the fluid flow at that point due to the other vortex line~\cite{Nozieres1999}. It is useful to express this interaction in terms of the frequency at which two co-rotating vortices orbit one another,
$\ifreq(r_{12}) =  {\hbar}/({mr_{12}^2)},$
where $m$ is the atomic mass and $r_{12}$ is their separation. The linear speed of a dipole is $v= r_{12}\ifreq(r_{12})$.

Each vortex in a ``gas'' of $n$ quantized vortices moves in response to the sum of two velocities,
\begin{equation}\label{eq:ode}
i\dot{z_k} = - S_k \pfreq(r_k) z_k + \frac{b}{2}\sum_{j\neq k}^n S_j \ifreq(r_{jk}) (z_k-z_j),
\end{equation}
where $z_k = x_k + i y_k=r_k e^{i \theta_k}$ is the complex coordinate of the $k$th vortex,
$S_k = \pm 1$ is its topological charge,
and
$r_{jk}=|z_k-z_j|$.
%, and the topological charge of the $k$th vortex is $S_k = \pm 1$. %, with the positive (negative) sign referring to counterclockwise (clockwise) circulation.
The constant parameter $b$ modifies the interaction strength slightly from the homogeneous case; for our experimental conditions, $b=1.35$~\cite{Middelkamp2010}. This lowest-order model does not take into account effects such as the bending of the vortex lines within the oblate condensate or the dissipative interactions that occur between vortex lines and thermal atoms at finite temperature~\cite{Jackson2009}.
%117

\begin{figure}[ht]
\includegraphics[width=8.5cm]{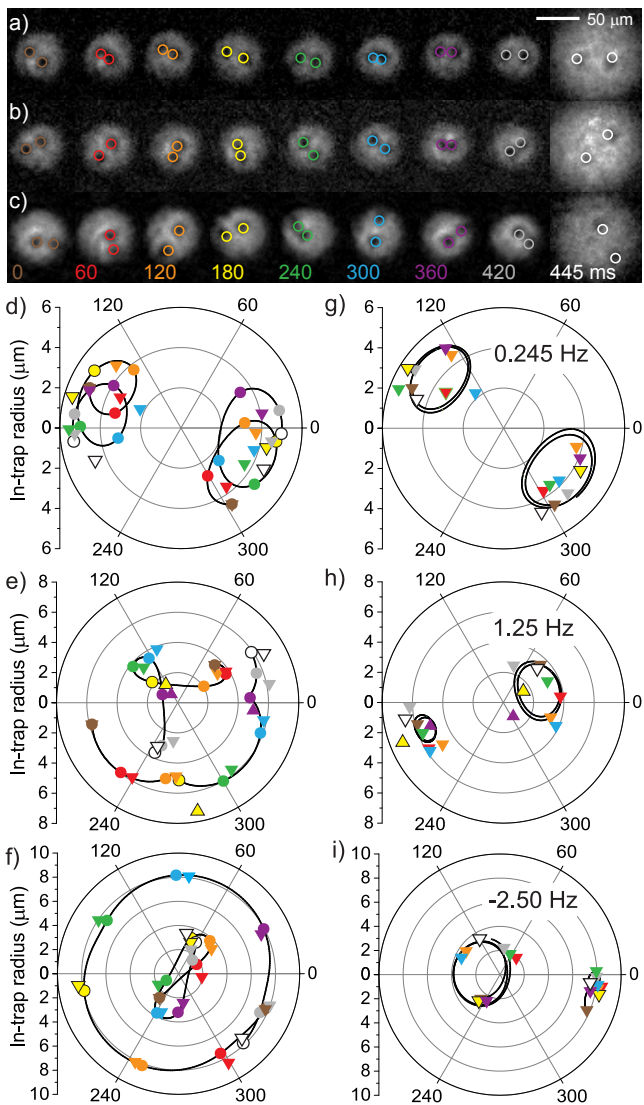}
\nuke{
\includegraphics[width=8.5cm]{translated1.eps}\\
%\hskip-3.5cm
\vspace{0.2cm}
\includegraphics[width=8.5cm]{translated2.eps}}
\caption{\label{fig:translated}
(color) Asymmetric vortex configurations. (a--c) Experimental images of the atomic density overlaid with calculated vortex positions from the model, \refeq{eq:ode} (colored rings). The images in (c) originally appeared in Ref.~\cite{Freilich2010}.
(d--f) Trajectories for the data in panels (a--c), using the notation of \reffig{fig:epicyclic}.
(g--i) Trajectories for the data in panels (d--f) in the indicated rotating frames. In (a--b) the extracted fraction is 1\%, whereas in (c) it is 5\% and the image intensity is scaled by a factor of two.
}
\end{figure}

{\it Analysis.---} The many vortex dynamics described by \refeq{eq:ode} provide a number of experimentally accessible predictions for the
observed symmetric and asymmetric vortex dipole configurations. First, an explicit prediction can be made about the stationary equilibrium positions of the constituent vortices. The velocities of the two vortices are zero for all times at a separation distance:
\begin{equation}\label{equil}
d_\mathrm{sep}=2 \rperp \sqrt{\frac{b\ifreq_R }{4\Omega_0 + b\ifreq_R}},
\end{equation}
with $\Omega_0\equiv\pfreq(0)$ and $\ifreq_R\equiv \ifreq(\rperp)$.
%This expression
Equation~(\ref{equil}) yields an average equilibrium distance of $0.436 \rperp$, in remarkable agreement with the experimental value $0.432(17)\rperp$~\cite{Freilich2010}.

The frequency of the epicyclic motion observed in \reffig{fig:epicyclic} is
calculated by linearizing the two-vortex equations around $d_\mathrm{sep}$. We obtain two pairs of frequencies, $\omega_{1,2}=0$, and
\begin{equation}\label{frequency1}
\omega_{3,4}=\pm \sqrt{2} \Omega_0 \left(1 + \frac{b\ifreq_R}{4 \Omega_0}\right)^{3/2}.
\end{equation}
The first pair illustrates the rotational invariance (around the center of the condensate) of the dipole~\cite{Pietila2006}, whereas the second pair corresponds to the
%so-called
anomalous modes of the vortex system \cite{Fetter2001}, similar to those of dark solitons~\cite{Weller2008,Theocharis2010,Coles2010}. These anomalous modes are responsible here for the counter-rotating, oscillatory vortex motion about the equilibrium points~\cite{Middelkamp2010}, which can be understood intuitively as the result of competition between the precession and interaction velocities~\cite{Li2008}. The measured oscillation frequency is $1.66(5)\,\overline{\Omega}_0$, where %$\overline{\Omega}_0$ is the
the overbar denotes the
average value during the experiment. This compares reasonably with the prediction of $1.52\,\overline{\Omega}_0$ from \refeq{frequency1}, %with the discrepancy due to
given the relatively large amplitude of the observed vortex motion. The initially cyclic vortex motion observed in Ref.~\cite{Neely2010} is also of this persuasion, albeit with larger amplitude.

A direct comparison between experimental trajectories and those predicted by the model is made possible by integrating the equations of motion (as shown in \reffig{fig:epicyclic}b). It is convenient to make this comparison by rescaling the post-expansion data by the ratio $\rperp/\rperp^*$, where $\rperp^*$ is the measured condensate radius after expansion, and $\rperp$ is the theoretical in-trap Thomas-Fermi radius calculated for a condensate of $N(t)$ atoms. The theoretical points are calculated from \refeq{eq:ode}, using values for the chemical potential that are also determined from $N(t)$, and are updated after each measurement in the subsequent time-evolution. We choose the initial positions of the modeled vortices to coincide with those of the experiment at $t=0$. We also include, as a sole free parameter, a small ($\lesssim 1\,\mu$m) offset (i.e., uncertainty) of the condensate center $(\delta x, \delta y)$ with respect to all of the measured data points. The offset is attributed to a technical effect arising from atomic micromotion in the time-averaged potential of the magnetic trap;
%and
similar shifts have been observed in condensates with single vortex lines~\cite{Freilich2010}. With these parameters and initial conditions, we find good agreement between the experiment and the results of the numerical integration (\reffig{fig:epicyclic}b) for symmetric displacements of the vortices from the equilibrium points. %121

Next, we consider vortex trajectories arising from \emph{asymmetric} displacements of the vortex lines.
Three typical image sequences are shown in \reffig{fig:translated}a--f. In \reffig{fig:translated}d, cyclic motion akin to that of \reffig{fig:epicyclic} is superposed upon a slow precession. With increasing asymmetric displacements, \reffig{fig:translated}e--f, one vortex moves in loops near the center of the condensate, while the other orbits closer to the periphery. The agreement between the simulation and the experiment ranges between fair (d) and quite good (e--f), with more pronounced discrepancies when the vortices are close together and difficult to resolve.

%Surprisingly,
Interestingly, all three data sets of \reffig{fig:translated} exhibit essentially identical epicyclic behavior, as is easily seen by transforming in each case to a suitable co-rotating reference frame (\reffig{fig:translated}g--i).
We therefore arrive at a unified and intuitive general understanding of the vortex dipole behavior: the vortex lines oscillate in phase about stable equilibrium points (or ``guiding centers'') that precess rigidly
about the center of the condensate. Both the precession frequency and the cyclic frequency (in the rotating frame) increase with displacement from the condensate center, and the amplitude of the outer vortex orbit becomes smaller than that of the inner vortex; this last fact is due to the faster precession of the outer vortex chiefly
influencing the co-rotating frame frequency. The stationary vortex dipole, as well as the oscillating dipole
in \reffig{fig:epicyclic}, are special cases of this general motion in which
guiding centers are stationary in the laboratory frame.

Such guiding center motions can be mathematically identified in the context
of the underlying model (\refeq{eq:ode}). Assuming rigidly rotating solutions $z_1(t) = r_1 \exp (i\omega t)$ and $z_2(t) = - r_2\exp(i\omega t)$, in terms of the fixed distances of the vortices from the condensate center $r_1$ and $r_2$ (in units of \rperp), we find the following expression for the precession frequency $\omega$:
\begin{equation}
\omega = \frac{1}{2}\left[\Omega(0) (\alpha-\beta) + \gamma
b_0 \left(\frac{r_1}{r_2}-\frac{r_2}{r_1}\right)\right].
\label{preq}
\end{equation}
where $b_0=\hbar b/(m R_{\perp}^2)$, $\alpha=1/(1-r_1^2)$, $\beta=1/(1-r_2^2)$
and $\gamma=(r_1+r_2)^{-2}/2$. A polynomial equation can also be
derived for $r_1$ (and $r_2$). A (nonlinear) analysis of
such equilibria  illustrates their stability, thus confirming
the generic quasi-periodic nature of the vortex motion.

{\it Conclusions \& Outlook.---} We have shown that a simple vortex-particle model reproduces the main features of vortex dipole behavior, as is experimentally observed in oblate Bose-Einstein condensates. The theoretical prediction of the separation distance between the stationary fixed points and the frequency of small oscillations around these equilibria are in good agreement with the experimental data. We ultimately understand the generic vortex behavior and its quasi-periodic character as cyclic oscillations about a broader class of effective equilibria consisting of rigidly rotating guiding centers. Stationary or periodic motions arise as
%identifiable
special cases within this set.

Extensions of this simple vortex-particle approach may be applied to larger ensembles of vortices, including co-rotating vortices that have not crystallized into a lattice. By improving the ability of the apparatus to image the extracted atomic samples, we can contemplate the intriguing possibility of detecting additional vortex interactions through departures from the model, such as the effects of bent vortex lines, reconnections, and other phenomena associated with quantum turbulence.

This work was supported by the NSF through grants PHY-0855475, DMS-0349023, DMS-0806762, by the M.E.C. of Spain through grant MTM2008-02502 and from the Alexander von Humboldt Foundation. D.S.H. acknowledges conversations with J.~R. Friedman, T.~K. Langin, and E. Altuntas.

%\bibliographystyle{apsrev}
%\bibliography{dipole}

%***
% Insert bibliography here

%***

\end{document}